\documentclass[prl,twocolumn]{revtex4-1}
\usepackage{amssymb}
\usepackage{amsmath}
\usepackage{graphicx}
\usepackage{color}
\usepackage[usenames,dvipsnames]{xcolor}
\newcommand{\private}[1]{}
\newcommand{\comment}[1]{}

\newcommand{\Tr}{{\rm Tr}}

\newcommand{\thedel}{\ensuremath{\delta(\eta^2\!+\!1)}}
\newcommand{\thedelp}{\ensuremath{\delta'(\eta^2\!+\!1)}}

\newcommand{\be}{\begin{equation}}
\newcommand{\ee}{\end{equation}}
\newcommand{\bea}{\begin{eqnarray}}
\newcommand{\eea}{\end{eqnarray}}

\newcommand{\ket}[1]{|#1\rangle}

\newcommand{\tsub}[1]{\langle #1 \rangle}

\makeatletter
\def\l@subsection#1#2{}
\def\l@subsubsection#1#2{}
\makeatother

\begin{document}

\title{A New Class of Particle in 2+1 Dimensions}

\author{Philip Schuster}
\email{pschuster@perimeterinstitute.ca}
\affiliation{Perimeter Institute for Theoretical Physics,
Ontario, Canada, N2L 2Y5 }

\author{Natalia Toro}
\email{ntoro@perimeterinstitute.ca}
\affiliation{Perimeter Institute for Theoretical Physics,
Ontario, Canada, N2L 2Y5 }
\date{\today}

\begin{abstract}
In two spatial dimensions, spin characterizes how particle states re-phase under changes of frame that leave their momentum and energy invariant. Massless particles can in principle have non-trivial spin in this sense, but all existing field theories only describe the trivial case. 
This letter presents a field theory for a massless particle with non-trivial physical spin. 
These particles are the 2+1-dimensional analogues of ``continuous-spin'' particles in 3+1 dimensions, but here they 
have only two real degrees of freedom, related by parity. They can be understood as massless generalizations of 
anyons, but are simpler in key respects. 
\end{abstract}

\maketitle

%\newpage
%%%%%%%%%%%%%%%%%%%%%%%%%%%
In two dimensions, particles of momentum $\vec k$
with Lorentz-invariant dispersion relations $E^2=m^2+\vec k^2$ (working with unit velocity) 
can be classified according to their spin. 
Physically, spin characterizes how a single-particle state behaves under changes 
of frame that leave its three-momentum invariant. 
For a massive particle, we can go to a frame where the particle is at rest, 
so that  rotations ${\cal R}(\theta)$ by an angle $\theta$ leave the energy-momentum $k=(m,0,0)$ invariant. 
Such a state  $\ket{\psi(k)}$ transforms as
\be
\ket{\psi(k)} \rightarrow e^{i\theta s}\ket{\psi(k)}. \label{eq:massiveTrans}
\ee
That is, the wave-function {\it re-phases} under momentum preserving 
changes of frame by an amount determined by the spin quantum number $s$.
Famously, in 2+1 dimensions, $s$ can take on any real value, 
%so that a rotation by $2\pi$ need not return the state to itself 
and the wave-function need not be single-valued --- this is the phenomenon of fractional spin, as described by 
anyons \cite{Leinaas:1977fm,Wilczek:1982wy}. 
Of course, the absolute phase of a wavefunction is not directly observable, 
but \eqref{eq:massiveTrans} is reflected in the properties of observable interactions. 

For massless particles with $E^2=\vec k^2$, the situation is different.
Rotations no longer preserve $\vec k$ in any frame, but an appropriate combined boost and rotation does.
For concreteness consider $(k^0,k^x,k^y) = (E,0,E)$.   
Then we find that $(R + K)$ generates changes of frame that leave $k^{\mu}$
invariant, where  $K$ generates boosts in the $x$-direction (transverse to $\vec k$), 
and $R$ is the counter-clockwise rotation in the spatial plane. 
Under large transformations ${\cal T}(\alpha)\equiv e^{i\alpha(R + K)}$,
such a particle state can transform as 
\be
\ket{\psi(k)} \rightarrow e^{i\alpha s}\ket{\psi(k)}, \label{eq:masslessTrans}
\ee
where $s$ is again a real number. Unlike the rotation, $(R+K)$ is a non-compact generator --- transformations ${\cal T}(\alpha)$ 
with arbitrarily large $\alpha$ never return to the identity operator, but rather correspond to increasingly 
large boosts in the $x$ direction, followed by a rotation that approaches $\pi/2$ at large $\alpha$ and a compensating boost in the $-y$ direction.  Thus, the wave-function can be single-valued for all $s$.  
Relatedly, the notion of fractional statistics under adiabatic exchange of point-localized anyons does not generalize to massless particles.  

Surprisingly, all familiar \emph{massless} field theories in 2+1 dimensions describe the spinless case $s=0$ --- even those where the field carries a Lorentz index.
For example, propagating modes of a Maxwell vector field $A^\mu$ can be written in Lorentz gauge with $k\cdot A=0$.   
The action of ${\cal T}(\alpha)$ on such a field simply maps $A^{\mu}\rightarrow A^\mu + i \alpha k^\mu$
--- a pure gauge transformation that does not alter the physical state, even by a phase.

This letter describes the quantum mechanics of massless degrees of freedom with non-trivial spin in the preceding sense.
To start, it is helpful to frame the concept of spin in a more precise and Lorentz-covariant form. 
With $J^{\mu\nu}$ the covariant rotation+boost generators and $P^\mu$ the translation generators, 
the Lorentz invariant quantity that characterizes spin is the pseudoscalar 
$W=\frac{1}{2} \epsilon_{\mu\nu\rho} J^{\mu\nu} P^\rho$.  This commutes with 
all rotations, boosts, and translations, so its eigenvalue is an invariant characteristic of the particle.  
The only other such invariant is the particle's mass.  
%One-particle states with definite momentum $k^{\mu}$ in 2+1-dimensional quantum systems 
%can be characterized by their transformation under Wigner's ``little group'' \cite{Wigner:1939cj}, which is 
%just the group of rotations and boosts that preserve $k^\mu$.
With $P^\mu \rightarrow k^\mu$, $W$ is also the \emph{unique} generator of Lorentz transformations 
that leave $k^\mu$ invariant --- it generates the ``little group'' of $k^{\mu}$, as formulated by Wigner \cite{Wigner:1939cj}.
For a mass-$m$ particle in its rest frame, $W=m J^{12}$ -- a rotation.  
For a massless particle, $W$ is no longer a pure rotation in any frame but rather a combination 
of a rotation and a boost transverse to ${\vec k}$, as considered in \eqref{eq:masslessTrans}.
We refer to a particle on which 
$W$ acts non-trivially as a \emph{panyon} --- a state on which \emph{all members} of the little group act. 
Going back to $k^\mu = (E,0,E)$, we have $W= (R+K) E$.
  
The covariant way of expressing the spin transformations \eqref{eq:massiveTrans} and \eqref{eq:masslessTrans}
together is, therefore, that the momentum-preserving change of frame $e^{i\kappa W(k)}$ transforms a state of \emph{any} momentum as
\be
\ket{\psi(k)} \rightarrow e^{i\kappa\rho}\ket{\psi(k)}, \label{eq:CovActionOnStates}
\ee
where $\rho=m s$ and $\kappa = \theta / m$ for a massive particle at rest (as in \eqref{eq:massiveTrans}), while $\rho = E s$ and $\kappa=\alpha / E$ for a massless particle (as in \eqref{eq:masslessTrans}).  The eigenvalue $\rho$ of $W$ characterizes the spin of \emph{physical states} of any mass,  
yet no theory of massless particles with non-zero $\rho$ is known.

\paragraph{Vector Superspace and the Panyon Action}
The remainder of this letter presents a free field theory whose elementary degrees of freedom are two panyons with non-zero ``spin'' eigenvalues $W|\psi_\pm\rangle =\pm \rho |\psi_\pm\rangle$, related to one another by parity.  The theory is a 2+1-dimensional version of the one in \cite{CSPnewfield}.  
Panyons cannot arise from any finite tensor action, because any rank-$n$ tensor field is annihilated by $W^{n+1}$, in conflict with the desired action of $W$ on states.  
Roughly speaking, we will need a gauge theory with an infinite tower of fields, in which all but two components are either pure gauge or non-dynamical.  Such constructions are familiar from the field theory for elementary anyons developed by Jackiw and Nair \cite{Jackiw:1990ka}.  
Like anyons, an infinite-component description of \emph{elementary} panyons does not preclude their appearance as composites in a theory built from a finite  number of familiar fields \cite{Wilczek:1982wy}.  

We introduce a bosonic field $\Psi(\eta^\mu,x^\mu)$ that depends not only on a space-time coordinate $x$, but on a new auxilliary ``coordinate'' $\eta^\mu$.  The dependence on $\eta$ is restricted to be smooth, so that we can Taylor-expand 
\be
\Psi(\eta,x) = \sum_n \eta^{\mu_1} \dots \eta^{\mu_n} {\psi^{(n)}(x)}_{\mu_1\dots \mu_n} , \label{eq:smooth}
%= \sum_n \eta^{n} \cdot \psi^{(n)}(x), \label{eq:smooth}
\ee
with coefficients $\psi^{(n)}(x)$ that are rank-$n$ tensor-valued functions.  
Under Lorentz transformations $\Psi(\eta,x) \rightarrow \Psi(\Lambda^{-1} \eta, \Lambda^{-1} x + a)$, the coefficients 
$\psi^{(n)}(x)$ transform covariantly, i.e. $\psi^{(0)}(x) \rightarrow \psi^{(0)}(\Lambda^{-1} x)$, $\psi^{(1)}_\mu(x) \rightarrow \Lambda_\mu^{\nu} \psi^{(1)}_\nu(\Lambda^{-1} x)$, etc.  Thus, the field content of $\Psi$ is entirely equivalent to an infinite tower of fields, but viewing it as a function over the auxilliary $\eta$-space will be both technically simpler and more illuminating in what follows.
As we will show, only two degrees of freedom (which can loosely be thought of as arising from $\psi^{(0)}$ and $\psi^{(1)}$) are dynamical; here, unlike in other known massless theories,
boosts will mix the states described by these fields. 

The length of an $\eta$-vector is unphysical, as it can be absorbed into changes of the tensor fields' normalization.  This motivates studying actions localized on a Lorentz-invariant surface in $\eta$.  The \emph{orientation} of $\eta$ along that surface, meanwhile, will be used to encode a panyon's spin.  
The action for a panyon is
\be
% \int \! d^3x[d^3\eta] \thedelp \tfrac{1}{2}(\partial_\mu \Psi(\eta,x))^2 +\tfrac{1}{4}  \thedel (\Delta\Psi)^2,\label{CSPaction1}
 \int \! d^3x[d^3\eta] \thedelp \tfrac{1}{2}(\partial_\mu \Psi)^2 +\tfrac{1}{4}  \thedel (\Delta\Psi)^2,\label{CSPaction1}
\ee
which is invariant under gauge transformations
\be
\delta\Psi_{\epsilon,\chi} = (\eta\cdot\partial_x - \tfrac{1}{2} (\eta^2+1) \Delta) \epsilon(\eta,x) + (\eta^2+1)^2 \chi(\eta,x),
\ee
where $\delta'(x) = \frac{d}{dx}\delta(x)$, $\Delta = \partial_\eta.\partial_x + \rho$, and $\epsilon$ and $\chi$ are, like $\Psi$, arbitrary smooth functions of $x$ and $\eta$.  
The brackets around $[d^3\eta]$ denote that this integral must be regulated.  For example, the contribution of the scalar component $\psi^{(0)}$ to the action involves the divergent integral $\int [d^3\eta] \thedelp$.  As discussed in \cite{CSPnewfield}, integrals of this form are entirely determined by their symmetry properties up to an overall normalization, or equivalently they can be defined by analytically continuing $\eta^0$.  The analytic continuation turns the integral over a hyperboloid in $\eta$ into a simple integral over a sphere.  Generating functions (following the normalization conventions of \cite{CSPnewfield})
\bea
G(w) &\equiv & \int [d^3 \eta] \thedel e^{i \eta.w} =  \frac{\sin(x)}{x} |_{x=\sqrt{-w^2}}\label{Gfunc}\\
G'(w) &\equiv &  \int [d^3\eta] \thedelp e^{i \eta.w} = \frac{\cos(x)}{2} |_{x=\sqrt{-w^2}}. \label{Gpfunc}
\eea
can be used to recast the action and other localized $\eta$-space integrals as operators acting on functions of $\eta$:
\be
\int [d^3 \eta] \thedel F(\eta) = \left[ G(i \partial_\eta) F(\eta)\right]_{\eta=0}, \label{FstarG}
\ee
and similarly for $\delta'$.   The expressions \eqref{FstarG} satisfy identities equivalent to integration by parts and $\thedel (\eta^2+1)=0$, so that manipulations in the tensor decomposition and in superspace are equivalent.

\paragraph{Physical Degrees of Freedom}
Varying \eqref{CSPaction1} with respect to $\Psi(\eta,x)$ we obtain the covariant equation of motion
\be
\delta^{\prime}(\eta^2+1)\Box_x \Psi - \tfrac{1}{2}\Delta(\thedel\Delta \Psi)=0 \label{eq:etaEOM}.
\ee
This equation simplifies when we transform to a convenient ``harmonic'' gauge where $\thedel\Delta \Psi = 0$.  This gauge can be reached from an arbitrary $\Psi$ by taking $\epsilon = - \frac{1}{\Box} \Delta \Psi$.  In this gauge, we have $\thedelp \Box_x \Psi = 0$, or equivalently $\Box_x \Psi = (\eta^2+1)^2 \beta(\eta,x)$ for an arbitrary smooth function $\beta$. 
A further transformation by $\chi =  -\frac{1}{\Box} \beta$ redues the equation of motion to 
\be
\Box_x \Psi = 0 \qquad \thedel\Delta \Psi=0 \label{eq:etaEOMharmonic}.
\ee
The above demonstrates that the excitations of the field $\Psi$ are massless.  
A residual gauge redunancy with $\Box_x\epsilon = \Box_x \chi = 0$ preserves this gauge choice.  

There are several ways to see that \eqref{eq:etaEOMharmonic} has exactly two gauge-inequivalent propagating solutions, notwithstanding the infinite tower of tensors in $\Psi$.

We first consider an arbitrary plane wave solution of 
 \eqref{eq:etaEOMharmonic}, which can be written as $\Psi_k(\eta,x) = e^{ik.x} \psi_k(\eta)$ where $k^2=0$ is null and $\thedel \Delta\psi_k(\eta) = 0$, or in other words $\Delta\psi_k(\eta) = (\eta^2+1)\alpha_k(\eta)$ for some smooth function $\alpha_k(\eta)$ (for momentum-space plane waves of momentum $k$, $\Delta = i\,k.\partial_\eta+\rho$).  The gauge parameters $\epsilon$ and $\chi$ that preserve the harmonic gauge \eqref{eq:etaEOMharmonic} can also be decomposed into plane waves; we can reach a gauge where 
\be
\Delta \psi_k(\eta) = 0,\label{strictGauge}
\ee
i.e. where the residual function $\alpha_k$ vanishes, by a gauge transformation satisfying $\Delta^2\epsilon_k(\eta) = 2 \alpha_k(\eta)$, which has solutions for any $\alpha_k$.

To fully fix gauge and identify the physical degrees of freedom, we introduce reference vectors $q$ and $\epsilon$ for each null momentum $k$, with $q.k = - \epsilon^2 = 1$ and $q^2=q.\epsilon=k.\epsilon=0$.  For example, we can take $q$ to be $(k^0,-{\vec k})/{|\vec k|}^2$ and $\epsilon$ to be the spatial vector for which $(\vec k,\epsilon)$ forms a right-handed coordinate system.  
One can fix a gauge in which $\psi_k(\eta) = e^{i\rho \eta.q} \sum_n c_n (\eta.\epsilon)^n$ (the logic closely follows \S IV.B of \cite{CSPnewfield}), with residual gauge freedoms of the form $\delta \psi_k = e^{i\rho \eta.q} [(\eta.\epsilon)^{n+2} - (\eta.\epsilon)^n]$ generated by ${\epsilon_n \propto e^{i\rho \eta.q} \eta.q (\eta.\epsilon)^n}$ for each $n\ge 0$ (where we have decomposed the metric as $g^{\mu\nu}\! = q^\mu k^\nu\! +\! k^\mu q^\nu\! -\! \epsilon^\mu \epsilon^\nu$).  
These gauge transformations relate all terms in $\psi_k(\eta)$ with even $n$ to one another, and similarly for odd $n$.  It follows that any plane wave of momentum $k$ satisfying \eqref{eq:etaEOMharmonic} can be decomposed into two physical modes $\psi_{k,0}(\eta) = e^{i\rho \eta.q} $ and $\psi_{k,1}(\eta) =\eta.\epsilon e^{i\rho \eta.q}$ plus pure gauge contributions.  

In the $\rho\rightarrow 0$ limit, these basis wavefunctions are precisely the scalar and vector components of $\Psi$.  The fact that we have only two propagating degrees of freedom, despite infinitely many tensors, is closely related to the fact that massless symmetric tensor gauge fields with rank $\ge 2$ have no propagating degrees of freedom in two spatial dimensions.  

To sharpen this connection, we can explicitly count degrees of freedom in the various tensor components $\psi^{(n)}(x)$ of \eqref{eq:smooth} in a covariant gauge, starting from \eqref{strictGauge}. 
We note first that for any function $\xi(\eta,x)$, the gauge variation of $\Psi$ generated by $\epsilon = (\eta^2\!+\!1)\xi(\eta,x)$ is precisely equivalent to the variation generated by $\chi = -\frac{1}{2} \Delta \xi$.  To avoid this double-counting, we should restrict $\epsilon$ (or $\xi$)  --- a convenient choice is $\partial_\eta^2 \epsilon =0$.  Choosing a gauge where $(\partial_\eta^2)^2 \Psi = 0$ fully fixes $\chi$.  Decomposing $\Psi(\eta,x)$ into  tensor components $\psi^{(n)}(x)$ as in \eqref{eq:smooth} (and similarly for $\epsilon(\eta,x)$), the above constraints amount to fixing $\psi^{(n)}$ to be double-traceless and $\epsilon^{(n)}$ to be traceless for all $n$.

It is further possible to reach a gauge where $\psi^{(n)}$ is traceless (i.e. in $\eta$-space, $\partial_\eta^2 \Psi = 0$) via a gauge transformation satisfying
\bea
%\delta_{\epsilon}(\partial_\eta^2 \Psi) & =& 
2\left(\partial_\eta.\partial_x - (\tfrac{3}{2}+\eta.\partial_\eta)\Delta \right) \epsilon  = -\partial_\eta^2 \Psi. \label{tracelessGaugeChoice}
\eea
In components, this takes the form $A_n \partial.\epsilon^{(n)}+ B_n \rho \epsilon^{(n-1)} = \partial_\eta^2 \psi^{(n)}$ where $A_n$ and $B_n$ are numerical coefficients. This equation is clearly invertible at each rank $n$.   
What is less obvious is that an $\epsilon$ that preserves $\partial_\eta^2 \Psi  =0$ (i.e. $\epsilon$ satisfies the homogeneous form of \eqref{tracelessGaugeChoice}) can also satisfy the $\Delta^2 \epsilon = 0$ condition required to keep $\Psi$ in the gauge \eqref{strictGauge}.  But in fact, 
by acting on \eqref{tracelessGaugeChoice} with $\Delta$ we can show that \eqref{tracelessGaugeChoice} \emph{guarantees} $\Delta^2 \epsilon = 0$.  

%% It is further possible to reach a gauge where $\psi^{(n)}$ is traceless (i.e. in $\eta$-space, $\partial_\eta^2 \Psi = 0$) by a suitable gauge transformation.  To see this, we note that 
%% \bea
%% \delta_{\epsilon}(\partial_\eta^2 \Psi) &= &[\partial_\eta^2, \eta.\partial_x - \tfrac{1}{2} (\eta^2+1) \Delta] \epsilon \nonumber \\
%% & =& 2\left(\partial_\eta.\partial_x -(\tfrac{3}{2}+\eta.\partial_\eta)\Delta \right) \epsilon. \label{tracelessGaugeChoice}
%% \eea
%% We can invert the above to find an $\epsilon$ that brings any $\Psi$ whose tensor components are double-traceless into a traceless gauge. This is easiest to see by expanding $\epsilon$ in components, so that $\eta.\partial_\eta$ is just counting the tensor rank of the objects to its right.  
%% What is less obvious is that an $\epsilon$ that preserves $\partial_\eta^2 \Psi  =0$ (i.e. where the expression \eqref{tracelessGaugeChoice} vanishes)  can also satisfy the $\Delta^2 \epsilon = 0$ condition required to keep $\Psi$ in the gauge \eqref{strictGauge}.  But in fact, 
%% if we act with $\Delta$ on \eqref{tracelessGaugeChoice}, we find that it \emph{guarantees} $\Delta^2 \epsilon = 0$.  

To summarize, for null momenta $k$ we can reach a gauge where $\Delta \psi_k = \partial_\eta^2 \psi_k = 0$, which is preserved by gauge transformations $\epsilon_k$ for which \eqref{tracelessGaugeChoice} vanishes and $\partial_\eta^2 \epsilon_k =0$.  
These conditions can be expressed in tensor form by the covariant gauge fixed equation of motion
\be
\Box \psi^{(n)}=0, \quad \Tr[\psi^{(n)}]=0 ,\quad  \partial\cdot \psi^{(n)}=\tfrac{\rho}{n}\psi^{(n-1)}, \label{conditions}\\
\ee
with residual gauge symmetry generated by $\epsilon^{(n)}$ satisfying $\Tr[\epsilon^{(n-1)}]=0$ and $\partial\cdot \epsilon^{(n-1)} = a_n \rho \epsilon^{(n-2)}$.
The coefficient $a_n$ follow from \eqref{tracelessGaugeChoice} but its value is unimportant.  
We can now count the number of unconstrained components in $\psi^{(n)}$, subtracting the number of unconstrained gauge parameters in $\epsilon^{(n-1)}$.  
For the scalar $n=0$,  all of the conditions above are trivial so the single component $\psi^{(0)}$ describes one degree of freedom.  
At $n=1$ the vector has three components, of which one is related to $\psi^{(0)}$ by the ``transverse'' condition, 
while one is pure gauge associated with the unconstrained $\epsilon^{(0)}$. This leaves one dynamical degree of freedom in $\psi^{(1)}$.
At $n=2$, the tensor has 6 components, of which one is removed by the trace condition and three are fixed by the ``transverse'' condition.  The remaining two degrees of freedom are pure gauge associated with the two non-transverse components of $\epsilon^{(1)}$. 
For higher ranks there continue to be no unconstrained components in $\Psi$.  
This counting is precisely the same as for familiar gauge theories.  In the case of $\rho=0$, only $\psi^{(0)}$ and $\psi^{(1)}$ are non-vanishing for the propagating modes.  For non-zero $\rho$, the basis functions $\psi_{k,0}$ and $\psi_{k,1}$ introduced earlier are particular solutions of the conditions \eqref{conditions}, which (as expected from \eqref{conditions}) have all tensor components non-zero.
%, with the residual gauge variations corresponding to changes of the pair of reference vectors ${q^\mu,\epsilon^\mu}$, accompanied by a rephasing.    
Of course, when $\rho=0$ higher-rank gauge fields can have important topological effects, even though they do not contain propagating modes; similar phenomena could be important in the present context.  

We can now check how the ``spin'' operator $W$ acts on the two basis modes.  
The Lorentz-invariant form of $W = i \epsilon_{\mu\nu\rho} \eta^\mu \partial_\eta^\nu k^\sigma$ can 
be re-expressed in terms of our reference vectors as $W= i(k\cdot \eta \, \epsilon \cdot \partial_\eta- \epsilon\cdot \eta \, k \cdot \partial_\eta)$, so that
\bea
W \psi_{k,0}(\eta) &=& \rho \psi_{k,1}(\eta) \label{eq:0mode} ,\\
W\psi_{k,1}(\eta) &=& e^{i\rho \eta.q}(-i\,k.\eta + \rho (\epsilon.\eta)^2) \simeq \rho \psi_{k,0}(\eta), \label{eq:1mode}
\eea
where $\simeq$ denotes identity up to a gauge transformation, generated in this case by $\epsilon = (-1+2i \eta.q) \psi_{k,0}$.  
The eigenmodes of $W$ at momentum $k$ are therefore ${\psi_{k,\pm} \equiv \psi_{k,0} \pm \psi_{k,1}}$, with eigenvalues $\pm \rho$  respectively.

This is all we need to carry out a general mode decomposition of the real field $\Psi(\eta,x)$ as
\bea
&& \Psi(\eta,x) = \nonumber \\
&& \int \frac{d^2k}{2|k|}\sum_{i=\pm}\left( a_{i}(k)\psi_{k,i}(\eta)e^{-ik\cdot x}+a^*_{i}(k)\psi_{k,i}^*(\eta)e^{ik\cdot x}\right). \nonumber 
%\label{eq:modeexpansion}
\eea
While a detailed discussion of the quantization of the action is beyond the scope of this short letter, 
we can sketch the final results. The mode coefficients $a_{\pm}(k)$ become annihilation operators 
for single particle states satisfying commutation relations
\be
[a_{\pm}(k),a_{\pm}(k')^*]=2|k|\delta^2(k-k'). \label{eq:CommRelations}
\ee
%
%which follows from covariant quantization of the action \eqref{CSPaction1} \cite{CSPnewfield,CSPjenny}. 
The action of $W$ follows directly from \eqref{eq:0mode} and \eqref{eq:1mode} as
\be
[W,a_{\pm}(k')^*] = \pm \rho a_{\pm}(k')^*,
\ee
and so single particle states $\ket{k,\pm}=a_{\pm}(k')^*\ket{0}$ satisfy
\be
e^{i\kappa W}\ket{k,\pm}=e^{i\kappa\rho}\ket{k,\pm},
\ee
as desired for a massless particle with spin. 
 
%When $\rho=0$, this formalism in fact reduces to familiar field theories. 
We can add current interactions of the form
\be
S_{int} = \int d^3 x [d^3 \eta]  \thedelp \Psi(\eta,x) J(\eta,x), \label{eq:source}
\ee
provided the current $J$ satisfies a ``continuity condition'' (for $\rho\neq 0$)
\be
\thedel (\partial_x.\partial_\eta + \rho) J = 0 
\ee
or, with an appropriate component expansion of $J(\eta,x)$
(so that \eqref{eq:source} takes the form 
${\sum_n (-1)^{n} \phi^{(n)}(x)\cdot J^{(n)}(x)}$ in components --- see \cite{CSPnewfield}),
\bea
\partial \cdot J_1 + \rho J_0 &=& 0 \nonumber \\
\partial \cdot J_2^{\mu} + \rho J_1^{\mu} &=& 0 \nonumber \\
\tsub{\partial \cdot J_3^{\mu\nu} + \rho J_2^{\mu\nu}} &=& 0 \quad ...etc\nonumber
\eea
where $\tsub{\dots}$ denotes the traceless part of the enclosed tensor.  
This reproduces the continuity conditions for familiar interactions when $\rho=0$.  
We do not yet know how this condition can be satisfied by an interacting theory with $\rho\neq 0$, but can gain some intuition from power-counting. If the scalar current $J_0(x)$ is non-zero, the first continuity condition 
requires a $J_1^{\mu}(x)$ at $O(\rho)$, with non-zero divergence, and the next continuity condition implies non-zero $J_2^{\mu\nu}(x)$ at $O(\rho^2)$, and so on.
Thus, the dominant interacting mode would be scalar-field-like.
In the absence of a $J_0$ coupling, 
we could instead start with an {\it conserved} vector current $J_1^{\mu}(x)$ at $O(\rho^0)$, in which case 
continuity implies an $O(\rho)$ $J_2^{\mu\nu}(x)$ with non-trivial divergence, and so on.
In this case we would expect the dominant interaction to be gauge-field-like.
This is a 2+1-dimensional version of the {\it helicity correspondence} discovered in \cite{CSP1,*CSP2,*CSP3}.

Symmetry arguments also shed light on some physical effects of non-zero $\rho$. 
For example, if a panyon has a 3-point interaction with a scalar matter particle, 
we can constrain the panyon emission and absorption amplitudes in the limit that the 
panyon momentum $k^\mu$ is much softer than the matter particle's momentum $p^{\mu}$, as in \cite{Weinberg:1964ew}.
In this limit, the soft factor appearing in 
scattering amplitudes where single $\ket{k,\pm}$ emission occurs must take the form
\be
S_{\pm}(p,k) \propto e^{\pm i\rho z},
\ee
where $z=\frac{\epsilon\cdot p}{k\cdot p}$.
This is the only function of $k$ and $p$ that transforms correctly as a single panyon state. 
Recall that the states $\ket{k,\pm}=\ket{k,0}\pm\ket{k,1}$, so the ``scalar'' and ``gauge-field'' soft factors are 
\be
S_{0}(p,k) \propto \cos \rho z ,\quad
S_{1}(p,k) \propto \sin \rho z.
\ee
In the $\rho\rightarrow 0$ limit, the $\psi_0$ mode dominates emission, with corrections at $O(\rho^2)$,
while emission amplitudes for the $\psi_1$ mode are $O(\rho)$. 
Such soft factors would lead to modified emission and absorption, relative to a massless scalar field,
in the forward and/or soft limits where $\rho z$ becomes large.

\paragraph{Discussion}
Of course, a re-phasing of states as in \eqref{eq:CovActionOnStates} is only physical in the presence of Lorentz-invariant interactions.  
The free field theory presented here is, however, a key prerequisite to understanding panyon interactions.
There are several reasons to further explore the physics of panyons. 
First, it would be remarkable if interacting panyons, like anyons, can arise as quasi-particles in effectively 2+1-dimensional physical systems.
Second, the similarity of panyons to anyons at the level of little group transformations --- both are invariantly characterized by arbitrary real spin eigenvalues
$W\ket{\psi}=\rho\ket{\psi}$ --- suggests potentially interesting relations between the two in field theory.  For example, perhaps panyons can arise  from the massless limit of appropriate anyon theories, or conversely anyons may appear in a broken phase of panyon gauge theories.
Finally, panyons are analogous to continuous-spin particles (CSPs) in 3+1 dimensions \cite{CSP1,*CSP2,*CSP3},  which 
are poorly understood but might have interesting phenomenological applications.  
Since panyons have only two degrees of freedom (or one, in parity non-invariant theories), compared to the infinite tower of dynamical states in 3+1 dimensions, they may offer a more theoretically tractable toy model for CSP dynamics.  
We anticipate that the theory of panyons presented here will facilitate progress answering these questions. 

\paragraph{Acknowledgements}
The authors thank Carlos Tamarit for suggesting the term panyon.
This research was supported in part by Perimeter Institute for Theoretical Physics. Research at Perimeter Institute is supported by the Government of Canada through Industry Canada and by the Province of Ontario through the Ministry of Research and Innovation.

%%%%%%%%%%%%
\bibliography{3dCSPletter}
%%%%%%%%%%%f%%%%%%%%%%%%
\end{document}